\documentclass[twocolumn,prb,floatfix,nobibnotes]{revtex4-2}
\usepackage[utf8]{inputenc}
\usepackage{graphicx}
\usepackage[caption=false]{subfig}
\usepackage[colorlinks=true,citecolor=blue]{hyperref}
\usepackage[version=4]{mhchem}

\newcommand{\bj}{{\boldsymbol{j}}}

\newcommand{\br}{{\boldsymbol{r}}}
\newcommand{\bbm}{{\boldsymbol{m}}}
\newcommand{\bn}{{\boldsymbol{n}}}
\newcommand{\bbM}{{\boldsymbol{M}}}
\newcommand{\bH}{{\boldsymbol{H}}}
\newcommand{\bB}{{\boldsymbol{B}}}
\newcommand{\bD}{{\boldsymbol{D}}}
\newcommand{\bE}{{\boldsymbol{E}}}
\newcommand{\bP}{{\boldsymbol{P}}}
\newcommand{\bM}{{\boldsymbol{M}}}

\newcommand{\bmu}{{\boldsymbol{\mu}}}
\newcommand{\bmd}{{\dot{\boldsymbol{m}}}}
\newcommand{\bnd}{{\dot{\boldsymbol{n}}}}

\newcommand{\ex}{{\hat{\boldsymbol{x}}}}
\newcommand{\ey}{{\hat{\boldsymbol{y}}}}
\newcommand{\ez}{{\hat{\boldsymbol{z}}}}

\begin{document}
\title{Impact of spin torques and spin pumping phenomena on magnon--plasmon polaritons in antiferromagnetic insulator--semiconductor heterostructures}
\author{Vemund Falch} 
\author{Jeroen Danon}
\author{Alireza Qaiumzadeh}
\author{Arne Brataas}
\affiliation{Center for Quantum Spintronics, Department of Physics, Norwegian University of Science and Technology, NO-7491 Trondheim, Norway}
\date{March 11, 2024}

\begin{abstract}
We investigate the impact of spin torque and spin pumping on the surface magnon polariton dispersion in a antiferromagnetic insulator--semiconductor heterostructure. In the bilayer system, the surface magnon polaritons conventionally couple to the plasma-oscillations in the semiconductor via electromagnetic fields. Additionally, magnons in the antiferromagnetic insulator layer may interact with the semiconductor layer via spin torques and their reciprocal phenomena of spin pumping. Due to the spin-to-charge conversion from the spin Hall and inverse spin Hall effects in the semiconductor layer with a strong spin-orbit coupling, this can couple the magnons to the plasmons in the semiconductor layer. Our research reveals that modifications in the mode frequency and the hybridization gap induced by these phenomena depend on the thickness of the antiferromagnetic layer. In thick layers, the spin-pumping contribution to the frequency shift and damping is inversely proportional to the wavelength, while in thin layers it is inversely proportional to the thickness. Furthermore, hybridization of the surface magnon polariton and dispersive magnons in the antiferromagnet is shown to depend on both the thickness and wavelength of the modes.

\end{abstract}

\maketitle

\section{\label{sec:Introduction}Introduction}

Magnonics and plasmonics are two active research fields in physics, photonics, and materials sciences. Plasmons, collective oscillations of the electron density, can miniaturize optical devices by generating and manipulating signals at optical frequencies along metal-dielectric interfaces on the nanometer scale, a dimension significantly smaller than the optical wavelength \cite{Maier:Springer2007}. These surface plasmons polaritons, confined at the conductor interface, allows for strong light-matter coupling with potential applications \cite{Poddubny:Nature2013,Uchida:NatComm2015,SPP_Magnon}. On the other hand, magnonics is a research field in magnetism that addresses the use of magnons, collective excitations of magnetic order, to transmit, store, and process information \cite{Chumak2015;MagnonicsReview,Magnonics1,Baumgaertl2023:Magnonics2}. The typical wavelength of magnons is orders of magnitude shorter than that of photons of the same frequency and thus makes it possible to build up more compact devices. 

Magneto-plasmonics represents the convergence of two distinct areas of research. The fundamental questions in this field are how magnons and plasmons, each with unique properties, can be effectively coupled to harness synergistic effects and how different system parameters can alter magneto-plasmonics modes. However, magnons and plasmons normally have very different energy scales, while ferromagnetic and antiferromagnetic (AF) systems have a characteristic energy scale of GHz and THz, respectively, three-dimensional (3D) metals host plasmons with an intrinsic optical gap \cite{VignaleBook}. Therefore, the interaction between magnons and plasmons is very weak and negligible. However, situations in two-dimensional materials and semiconductors can be different. Recent theoretical studies have shown the possibility of strong coherent coupling between 2D magnetic layers and plasmons \cite{PhysRevB.108.045414,PhysRevB.107.195302,Costa_2023}. The possibility of a strong interaction between plasmons, electromagnetic waves, and magnons in ferromagnetic and AF semiconductors with very low carrier density has also been pointed out decades ago \cite{PhysRev.131.512,Baryakhtar-FM,Baryakhtar-AFM,Baskaran}.

Antiferromagnets have recently attracted considerable interest \cite{Jungwirth:NNano2016,Gomonay:NPHYS2018,Baltz:RMP2018}. They have potential in THz electronics. Precessing spins in AF systems pump spin angular momentum into adjacent conductors \cite{Tserkovnyak:PRL2002,Brataas:PRB2002,Tserkovnyak:RMP2005,Cheng:PRL2014} that give rise to electric signals via the inverse spin Hall effect~\cite{Li:Nature2020,Vaidya:SCIENCE2020}. Electrical injection and detection reveal micrometer distance spin transport \cite{Lebrun:NATURE2018,lebrun_long-distance_2020,PhysRevB.107.184404}. 

Similarly to conducting interfaces, AF systems can host surface magnon polaritons (SMP) at their interfaces with other materials \cite{Camley:PRB1982,Jensen:PRL1995}, whose properties depend on the orientation and width of the AF layer \cite{Hale:PRB2022, Hao:IOP2023, Hao:PRB2021}. Several hybrid structures have also been investigated, which could be used to further modify the properties of SMP, as well as host several hybrid polariton modes \cite{Hale:PRB2022,Bludov:2DMat2019,Quang:PRM2022,Tarkhanyan:JMM2010}. SMPs could be used to create materials with negative refraction \cite{Macedo:PRB2013} and have been shown to enhance spin relaxation in nearby emitters \cite{Sloan:PRB2019}. More exotic states in AF structures have also been studied, Ghost SMPs that both oscillate and decay away from the interface \cite{Song:PRB2022} whose phononic counterpart has been observed experimentally \cite{Ma:Nature2021}. Dyakonov SMPs lie outside the AF Restrahlen band, where the permeability is negative and are instead carried by the anisotropic nature of AFs, which have also been predicted \cite{Hao:PRB2021}.

Recently, the coupling between plasmons in 2D graphene and magnons in uniaxial AF insulator (AFI) \cite{Bludov:2DMat2019,Pikalov} has been studied. The hybrid modes, surface magnon-plasmon polaritons, and surface plasmon-magnon polaritons, resulted from the two subsystems interacting via light. These hybrid excitations have intriguing properties that are controllable in new ways. For instance, the dispersion of the surface magnon--plasmon differs from that of uncoupled surface magnons, expressed in a reflection of the group velocity. Other recent related work investigates surface plasmon-phonon-magnon polaritons in topological insulators in contact with AFI \cite{Quang:PRM2022}, as well as the TE-surface plasmons in a 2DEG-(anti)ferromagnetic system \cite{yuan2024breaking}. 

\begin{figure}[t]
    \centering
\includegraphics[width=8.6cm]{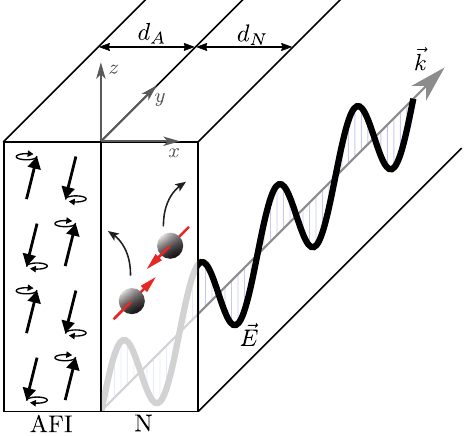}
    \caption{
    The geometry of the system, composed of a bilayer of an antiferromagnetic insulator of width $d_A$ and a normal semiconductor of width  $d_N$, is considered semi-infinite in the $yz$-plane and bounded by vacuum along the $x$-direction. The electromagnetic fields of the surface magnon polariton propagates along the interface, in the figure only the electric field $\boldsymbol{E}$ is shown. The spins in the AF and electrons in the N interact with each other and the electromagnetic fields, determining the behavior of the polariton.
    }
    \label{fig:SysGeom}
\end{figure}

At interfaces between magnets and conductors, spin transfer and spin-orbit torques and their reciprocal phenomena spin-pumping and charge-pumping couple the spin-dynamics in the magnets even when they are insulating and the spin and charge flow in the metals \cite{Ralph:JMMM2008,Brataas:NMAT2012,Tserkovnyak:RMP2005,Manchon:RMP2019,Brataas:PREP2020}. Furthermore, in conductors with spin-orbit interactions, the spin Hall effect couple spin and charge currents. We will explore how this coupling can alter surface magnon--plasmon polaritons. To this end, we consider a bilayer containing an AFI layer placed on top of a normal semiconductor (N) layer, as shown in Fig. \ref{fig:SysGeom}. In this geometry, magnons reside in the AFI layer, and surface magnon polaritons exist at the interfaces. In this way, the magnons have a long relaxation time since there are no lossy electrons to dissipate energy. In the absence of the spin-charge conversion and interfacial spin-mixing, the solutions to Maxwell's equations with the dielectric function of the conducting media and magnetic permeability tensor of the insulating magnet are known \cite{Hale:PRB2022, Maier:Springer2007}. The behavior of magnons pumping spins into a conducting medium with spin-charge conversion has also been studied \cite{Jiao:SpinCurrents}. Our challenge is to combine these descriptions and explore how the presence of interfacial spin-conductance and spin-charge conversion modifies the surface polaritons.

The rest of the paper is structured as follows: First, we present the system setup and the relevant equations governing the behavior of electromagnetic fields and charge- and spin-currents in Sec. \ref{sec:Model}. We then investigate the different modes that can exist in the bulk of the different materials in Sec. \ref{sec:bulk} before finding the dispersion relation of the SMPs by enforcing the boundary conditions in Sec. \ref{sec:bcs}. We investigate the effect of a finite exchange-stiffness, spin-pumping, and spin-Hall effect on the dispersion relation by both numerical and analytical means in section \ref{sec:res}, using a perturbative approach to derive the lowest-order frequency shifts in both the wide and narrow limit. We also investigate the hybridization between SMPs and bulk magnons.  We present our conclusions in section \ref{sec:Conclusion}.

\section{\label{sec:Model}Model}

\begin{figure}[t]
    \centering
    \includegraphics[width=8.6cm]{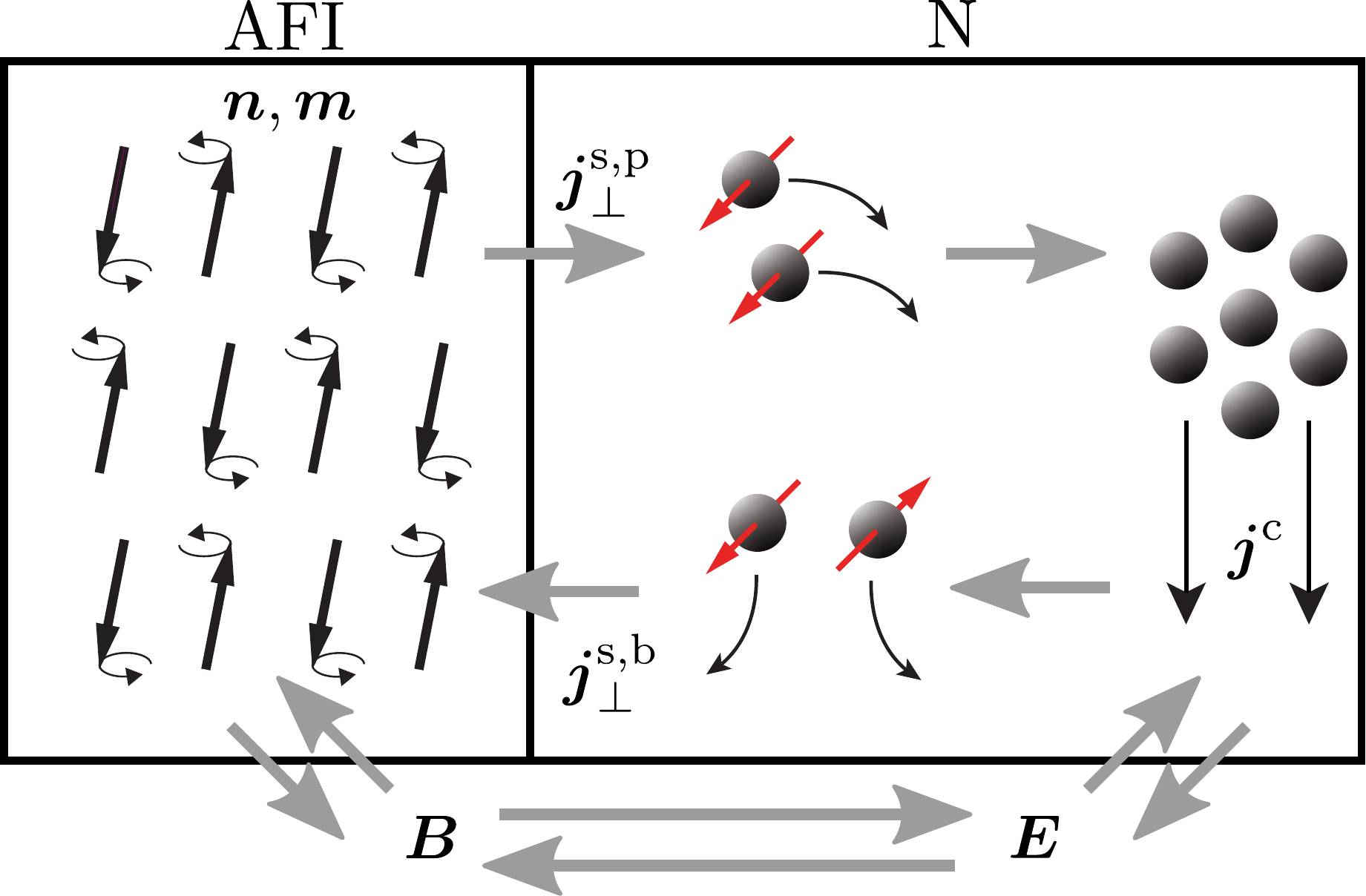}
    \caption{The coupling of the different modes in the system. The electromagnetic fields interact with the spins in the AFI layer, which are characterized by the orientation of the staggered $\bn$ and magnetization $\bbm$ fields, and with the electrons in the N layer where the interaction gives rise to a charge current density $\bj^c$.
    The resulting spin dynamics in the AF can pump spins into the N, giving rise to a diffusive spin current density $\bj_\perp^{s,p}$, which contributes to the charge current through the inverse spin-Hall effect.
    The charge current in the N also creates a spin accumulation in the N through the spin-Hall effect, which in turn contributes a spin current density $\bj_\perp^{s,b}$ that can cross the interface into the AF, where it influences the spin dynamics.}
    \label{fig:SysCoupl}
\end{figure}

We consider a system composed of an AFI-N bilayer, semi-infinite in the $y-z$ plane but finite along the $x$-direction, as shown in Fig.~\ref{fig:SysGeom}. A surface polariton can propagate at the interface between the two materials. As shown in Fig.~\ref{fig:SysCoupl}, the spin dynamics in the AFI layer and spin and charge dynamics in the N layer are coupled both through the electromagnetic fields and spin-Hall effect, as well as through the spin-pumping and spin-backflow through the AFI-N interface. 

In the absence of spin pumping and torque phenomena, the AFI and N layers interact through electromagnetic fields, as indicated in Fig.~\ref{fig:SysCoupl}, which are governed by the Maxwell equations.
These equations describe the coupled behavior of the electric $\bE$  and magnetic $\bH$  fields in all regions, including the vacuum on the two sides of the bilayer.
In the absence of free charges and currents, the Maxwell equations then read as,
\begin{subequations}
\begin{align}
	\begin{split}
		&\nabla\cdot\bD=0 , 
	\end{split}\label{eq:MaxwellDivE}\\
	\begin{split}
		&\nabla\cdot\bB=0 , 
	\end{split}\label{eq:MaxwellDivB}\\
	\begin{split}
		&\nabla\times\bE=-\frac{\partial\bB}{\partial t} , 
	\end{split}\label{eq:MaxwellCurlE}\\
	\begin{split}
		&\nabla\times\bH=\frac{\partial\bD}{\partial t},
\end{split}\label{eq:MaxwellCurlB}
\end{align}\label{eq:Maxwell}%
\end{subequations}
where 
\begin{equation}\label{polar}
\bD=\epsilon_0\bE+\bP,
\end{equation}
is the displacement field, with $\bP$ is the electric polarization of the material and $\epsilon_0$ is the vacuum dielectric constant, and 
\begin{equation}\label{mpolar}
\bH=\bB/\mu_0-\bbM,
\end{equation}
where $\bbM$ and $\bB$ are the magnetization of the system and magnetic flux density, respectively, and $\mu_0$ is the vacuum permeability.

\emph{AFI layer--} The Landau--Lifshitz--Gilbert (LLG) equation governs the dynamics of the spins in the AFI layer. For simplicity, we neglect the Gilbert damping parameter in our model. We are interested in the real part of frequency, and for the effects in question, the Gilbert damping can be neglected as a higher-order term. In a two-sublattice collinear AFI, the magnetic moments at the sublattice $a$ ($b$) are characterized by a unit vector $\bbm_{a (b)}$, and the coupled LLG equations can be expressed in terms of the magnetization $\bbm(\br,t)=\big(\bbm_a+\bbm_b)/2$ and staggered $\bn(\br,t)=\big(\bbm_a-\bbm_b)/2$ fields~\cite{PhysRevLett.106.107206},
\begin{subequations}
\begin{align}
    \bmd(\br,t)=&-\frac{\gamma}{2}\Big(\bbm\times \bH_\bbm+\bn\times\bH_\bn\Big), \label{LLGm}\\
    \bnd(\br,t)=&-\frac{\gamma}{2}\Big(\bbm\times \bH_\bn+\bn\times\bH_\bbm\Big)
    \label{LLGn} ,
\end{align}\label{eq:LLG}%
\end{subequations}
where $\gamma$ is the gyromagnetic ratio and the effective magnetic field $\bH_\bbm=-(1/M_s) \delta \mathcal{U} / \delta \bbm $ and staggered field $\bH_\bn=-(1/M_s) \delta \mathcal{U} / \delta \bn$ follow from the total potential energy $\mathcal{U}$, where $M_s$ denotes the sublattice saturation magnetization.  

We consider a uniaxial AFI layer with the following potential energy density within the exchange approximation \cite{AF_Free_Energy},
\begin{align}
    \begin{split}
    u=\frac{M_s}{\gamma}\Big[&\omega_\text{ex}\bbm^2+a\big(\nabla\bn\big)^2-2\mu_0\gamma\bH\cdot\bbm
    -\omega_a n_z^2\Big] ,
    \end{split}\label{eq:AFFreeEnergy}
\end{align}
where $\omega_\text{ex}$ is the exchange frequency, $a$ is the exchange stiffness, and $\omega_a$ the anisotropy energy. Using the procedure of Ref.~\cite{Rado:SpinCurrent}, it follows from the LLG equation \eqref{eq:LLG} that the spin current density in the AFI, within the exchange approximation, is given by,
\begin{equation}
    j_{i,\alpha}^{s,A}=-\frac{2 a M_s}{\gamma}\big(\bn\times\partial_i\bn\big)_\alpha,
    \label{eq:SpinCurrAF}
\end{equation}
where the first subscript $i$ denotes the direction of the current and the second index $\alpha$ relates to the orientation of the spins. We will use the sub- or superscript $A$ to denote quantities in the AFI and similarly $N$ for quantities in the normal semiconductor.

\emph{N layer--} The electric polarization of the semiconductor, Eq. (\ref{polar}), can be divided into two parts, one part $\bP_\infty$ coming from the core electrons and one part $\bP_c$ from the conduction electrons. The response of the former can be assumed to be proportional to the electric field and frequency-independent in the frequency region of interest, such that we can define $\bD=\epsilon_\infty\epsilon_0 \bE+\bP_c$, where $\epsilon_\infty$ is the high-frequency dielectric constant of the ion background. The time evolution of $\bP_c$ is governed by the differential equation $\partial_t \bP_c=\bj^c$, where the charge current density $\bj^{c}$ generally arises from the electrical conductivity of the material $\bj^{c,0}=\sigma\bE$ and the spin-to-charge conversion through the inverse spin-Hall effect $\bj^{c,{\rm SH}}$, see below for more details. 
For a 3D electron gas, the Drude conductivity at a given frequency $\omega$ reads,   
\begin{equation}
    \sigma(\omega)=\frac{\epsilon_0\omega_p^2\tau}{1-i\omega\tau} , 
\end{equation}
where $\tau$ is the elastic scattering time and $\omega_p$ the plasma frequency.

A spin current in the N layer leads to spin accumulation $\bmu^s$, which is governed by the following spin-diffusion equation \cite{Jiao:SpinCurrents}
\begin{align}
\begin{split}
    \frac{\partial \bmu^s}{\partial t}= {} & {} \gamma_N\bH\times \bmu^s-\Big(\frac{\hbar}{2}\nu\Big)^{-1}\partial_i j_{i,\alpha}^{s, N}\hat{\boldsymbol{e}}_\alpha-\frac{\bmu^s}{\tau_\text{sf}},
\end{split}\label{eq:SpinDiffusionEq}
\end{align}
where $\gamma_N$ is the gyromagnetic ratio in N, $\nu$ is the density of states per spin at the Fermi level, $\tau_\text{sf}$ is the spin-flip relaxation time, $j_{i,j}^{s,N}$ is the spin current density in the N layer, $\hat{\boldsymbol{e}}_\alpha$ is the unit vector in the $\alpha$ direction, and summation over repeated indices is implied. Since we do not consider an externally applied magnetic field, the term $\bH\times \bmu^s$ can be neglected as a second-order contribution in the small fields. The spin accumulation, in turn, gives rise to a spin current density,
\begin{equation}
	j^{s,0}_{i,\alpha}=- D_0\nu\frac{\hbar}{2}\frac{\partial\mu^{s}_{\alpha}}{\partial r_i},\label{eq:SpinDispersionCurrent}
\end{equation}
where $D_0$ is the diffusion coefficient of the N layer.

The charge and spin currents in the N layer couple via the spin-Hall and inverse spin-Hall phenomena~\cite{SHE1, SHE2, SHE3}, and we thus write for the total spin and charge currents in the N layer,
\begin{subequations}
\begin{align}
j^{s,N}_{i,\alpha}=j^{s,0}_{i,\alpha}+j^{s,\text{SH}}_{i,\alpha}&=j^{s,0}_{i,\alpha}-\theta_\text{SH}\frac{\hbar}{2e}\epsilon_{i\alpha j}j^{c,0}_{j}\label{eq:SHE},\\
j^{c}_i=j^{c,0}_i+j^{c,\text{SH}}_i&=j^{c,0}_i+\theta_\text{SH}\frac{2e}{\hbar}\epsilon_{ij\alpha}j^{s,0}_{j,\alpha}\label{eq:SpinHallCharge},
\end{align}
\end{subequations}
where $\epsilon_{ijk}$ is the Levi--Civita tensor and the spin-Hall angle $\theta_\text{SH}$, given as the ratio between the spin Hall and the charge conductivity, is assumed to be small. Although $\theta_\text{SH}$, in general, exhibits a non-trivial frequency dependence \cite{SHEFreq1, SHEFreq2}, we will treat it here as a constant, since SMPs only exist in a narrow frequency band close to the antiferromagnetic resonance frequency. This frequency region, where the real part of the permeability is negative, is known as the Reststrahlen band due to its reflective properties \cite{Hale:PRB2022}.

\emph{Boundary conditions --}
To solve the dynamics of the coupled bilayer system, we need the corresponding set of boundary conditions. The electromagnetic fields at the interface between two media $1$ and $2$ must satisfy,
\begin{subequations}
\begin{align}
	&\hat{\boldsymbol{e}}_{12}\cdot\big(\bD^{(2)}-\bD^{(1)}\big)=0 , 
	\label{eq:MaxwellBC_E_Orth}\\
	&\hat{\boldsymbol{e}}_{12}\cdot\big(\bB^{(2)}-\bB^{(1)}\big)=0 , 
	\label{eq:MaxwellBC_B_Orth}\\
	&\hat{\boldsymbol{e}}_{12}\times(\bE^{(2)}-\bE^{(1)})=0 , 
	\label{eq:MaxwellBC_E_Par}\\
	&\hat{\boldsymbol{e}}_{12}\times(\bH^{(2)}-\bH^{(1)})=0 , 
	\label{eq:MaxwellBC_H_Par}
\end{align}\label{eq:MaxwellBC}%
\end{subequations}
where the unit vector $\hat{\boldsymbol{e}}_{12}$ points from medium $1$ to $2$ perpendicularly to the interface.
%and $\sigma^c$ and $\boldsymbol{K}$ are the surface charge and (charge) current densities, respectively.
We consider an AFI with a compensated interface and thus ignore any net magnetization at the interfaces.
Furthermore, as all the interfaces are assumed to be clean and lacking spin-flip scattering, the conservation of spin angular momentum requires the spin current densities to be continuous across all the interfaces. This introduces an additional coupling channel between the dynamics in the AFI and N layers, as shown in Fig.~\ref{fig:SysCoupl}. The dynamics of the interfacial magnetic moments in the AFI layer give rise to a pumped spin current density into the N layer \cite{Cheng:PRL2014,TangCheng:InterSubLatticeGr}. Under the assumption that the cross-sublattice contribution is small \cite{TangCheng:InterSubLatticeGr}, the dominant contribution in the exchange limit to linear order is
\begin{equation}
    j^{s,p}_{\perp,\alpha}=\frac{\hbar g_r}{4\pi}\left(\bn\times\bnd\right)_\alpha, \label{eq:SpinPumping}
\end{equation}
where $g_{r}$ denotes the real part of the intralattice spin-mixing conductance per area. In addition to being small due to the exchange limit, the imaginary part of the spin-mixing conductance is often tiny compared to the real part \cite{Gi1,Gi2}.
A positive current $\bj^{s,p}_{\perp}$ is defined to flow in the positive $x$ direction, i.e., out of the AFI layer. The spin accumulation at the interface can also diffuse back into the AFI layer, which, under the same approximations as above, gives rise to a back-flow spin current density 
\begin{align}
    \begin{split}
        \bj^{s,b}_{\perp,\alpha}=&-\frac{g_r}{4\pi}\left(\bn\times\bmu^s\times\bn\right)_\alpha,
    \end{split}\label{eq:SpinBackflow}
\end{align}
across the interface. This leads to a torque acting on the magnetic moments in the AFI. The total spin current density crossing the interface is thus $\bj^{s,p}_\perp+\bj^{s,b}_\perp$. 

The boundary conditions for the LLG and spin-diffusion equations are given by (i) continuity of spin current densities at the AFI--N interface,
\begin{equation}%\bj_\perp^{s,A}=\bj^{s,p}_\perp+\bj^{s,b}_\perp=\bj_\perp^{s,S}\label{eq:BCAll}.
j_{x,\alpha}^{s,A}(x=0) = j^{s,p}_{\perp,\alpha}+j^{s,b}_{\perp,\alpha} = j_{x,\alpha}^{s,N}(x=0)\label{eq:BCAll},
\end{equation}
and (ii) vanishing of the spin current density at the AFI--vacuum and the N--vacuum interfaces,
\begin{equation}
    j_{x,\alpha}^{s,A}(x=-d_A) = j_{x,\alpha}^{s,N}(x=d_N) = 0.
    \label{eq:BCvac}
\end{equation}
Note that no similar boundary condition exists for the electric current, which can build up as a surface charge at the interfaces.

Solving these coupled equations for the charge and spin potentials in the N layer and the magnetic moments in the AFI layer, we can determine the magnon--plasmon polariton resonances and their properties.

\section{Bulk Behavior}\label{sec:bulk}
We start by considering the bulk behavior of the different layers in the system before imposing the boundary conditions between them.
For simplicity, we consider a case in which the propagation of the electromagnetic waves at the interface, the $\ey-\ez$ plane, is along the $\ey$ direction and the ground-state Néel vector direction is $\bn \parallel \ez$ in the AFI layer, see Fig. \ref{fig:SysGeom}.
This choice results in a relatively compact set of Maxwell equations to consider:
(i) Considering that the bilayer structure is still translationally invariant along the $\ez$ direction, we can assume that all fields are uniform in that direction; this causes the Maxwell equations \eqref{eq:MaxwellCurlE} and \eqref{eq:MaxwellCurlB} to decouple into one transverse electric (TE) and one transverse magnetic (TM) mode~\cite{Maier:Springer2007}.
(ii) The TM mode has $\bH=H_z\ez$, i.e., the magnetic field is always (anti)parallel to the ground-state Néel vector direction in the AFI layer and it thus does not couple to the spin dynamics to linear order; therefore, the TM mode is not able to host SMPs, and we will not consider it here.
We are thus left with the TE mode, with $\bE=E_z\ez$, causing the Maxwell equations to reduce to
\begin{subequations}\label{eq:MaxwellTE}
\begin{align}
&\partial_yE_z=-\partial_t B_x,
\\&\partial_xE_z=\partial_t B_y,\label{eq:TE_SpatE}\\
&\partial_xH_y-\partial_yH_x=\partial_t D_z.
\end{align}
\end{subequations}
We begin by finding the normal modes for this TE mode in the vacuum, the bulk AFI, and the bulk N. Due to the translational and temporal invariance of the bulk and the fact that all the equations will be linearized, we work in the Fourier space such that all fields have a simple exponential dependence $\propto e^{i(k y-\omega t)+qx}$. Note that the fields here refer to both electric $\bE$ and magnetic $\bH$ fields as well as the internal fields $\bm,\delta\bn=\bn-\ez$ and $\bbm$ in the AFI and $\bmu^s$ in N layers.
The wave number along the $\ey$ direction, $k$, and the frequency, $\omega$, are assumed to be given for now, and we want to find the wave number $q(k,\omega)$, which describes the localization of the mode along the $x$ direction. In general, $q$ is a complex number, becoming purely real or imaginary in limiting cases.

\emph{Bulk vacuum--}
In the vacuum, the bulk behavior is fully determined by the Maxwell equations \eqref{eq:MaxwellTE}, which can be rewritten in Fourier space as,
\begin{equation}
    \begin{bmatrix}
    -i \mu _0 \omega  & 0 & i k \\
 0 & -i \mu _0 \omega  & -q \\
 -i k & q & i \epsilon_0\omega
    \end{bmatrix}\begin{bmatrix}
        H_x\\H_y\\E_z
    \end{bmatrix}=0.
    \label{eq:Vacuum_qMat}
\end{equation}
Equation \eqref{eq:Vacuum_qMat} has solutions for 
\begin{equation}
    q_0^2=k^2-\frac{\omega^2}{c^2}\label{eq:qVacSol},
\end{equation}
where $c=(\epsilon_0\mu_0)^{-1/2}$ is the speed of light.
As we assume the vacuum to be semiinfinite on each side of the structure, we must have $\Re\{q_0\}\neq 0$ and thus $k>\omega/c$, so that the fields can vanish at $x\to\pm\infty$ and the wave is localized.

\emph{Bulk AFI--}
In the AFI, the bulk behavior is governed by the Maxwell equations \eqref{eq:MaxwellTE} and the coupled LLG equations \eqref{eq:LLG}. Solving the LLG equations for $\bbm$ and $\bn$ fields as a function of $\bH$ gives $\bM=2M_s\bbm=\chi_M\bH$ and  $\delta \bn=(\chi_N/M_s)\ez\times\bH$ with the following magnetic and staggered field response functions, respectively,
\begin{subequations}
\begin{align}
    \chi_M&=\frac{2\omega_s^2\big[1+\lambda_a^2(k^2-q^2)\big]}{\omega_0^2\big[1+\lambda_a^2(k^2-q^2)\big]-\omega^2},\\
    \chi_N&=\frac{i\omega}{\omega_a}\frac{\omega_s^2}{\omega_0^2\big[1+\lambda_a^2(k^2-q^2)\big]-\omega^2},
\end{align}\label{eq:SusceptibilitiesAF}%
\end{subequations}
where we have introduced the AF resonance frequency $\omega_0=(\omega_a(\omega_\text{ex}+\omega_a))^{1/2}$, the saturation frequency $\omega_s^2=\gamma\mu_0\omega_aM_s$, and the domain-wall length $\lambda_a^2=a/\omega_a$.
Using that $\bM=2M_s\bbm$, and substituting for $\bB=\mu_0(\bH+\bM)$, the Maxwell equations \eqref{eq:MaxwellTE} take the form
\begin{equation}
    \begin{bmatrix}
    -i \mu _0 \mu_m \omega  & 0 & i k \\
 0 & -i \mu _0 \mu_m \omega  & -q \\
 -i k & q & i \epsilon_0\omega
    \end{bmatrix}\begin{bmatrix}
        H_x\\H_y\\E_z
    \end{bmatrix}=0,\label{eq:AF_qMat}
\end{equation}
where $\mu_m=1+2M_s\chi_m$. This set of equations has six different solutions for $q$. The four modes with
\begin{align}
    q_{\text{A}\pm}^2= {} & {} k^2-\frac{1}{\lambda_a^2}\frac{\omega^2-\omega_0^2}{2\omega_0^2}-\frac{\omega^2}{c^2}\frac{\omega_0^2+2\omega_s^2}{2\omega_0^2}\label{eq:qAF_Hyb}\\
   &\pm\sqrt{\bigg(\frac{\omega^2-\omega_0^2}{2\lambda_a^2\omega_0^2}-\frac{\omega^2}{c^2}\frac{\omega_0^2+2\omega_s^2}{2\omega_0^2}\bigg)^2+\frac{2\omega^4\omega_s^2}{c^2\lambda_a^2\omega_0^4}},\nonumber
\end{align}
are hybrid modes containing a polariton and a magnon contribution, and can be found by replacing $\mu_0\to\mu_0\mu_m$, or equivalently $c^{-2}=\mu_0\epsilon_0\to c^{-2}\mu_m$, in Eq.~\eqref{eq:qVacSol} and solving for $q$.
The other two modes have
\begin{align}
    &q_{\text{A}0}^2=k^2+\frac{1}{\lambda_a^2} \left(1-\frac{\omega ^2}{\omega_0^2+2
   \omega_s^2}\right),\label{eq:qa0}
\end{align}
yielding $\mu_{m}=0$ for all $k,\omega$, which can be seen to make the matrix in Eq.~\eqref{eq:AF_qMat} non-invertible. These two modes have no electric field components and represent a correction to the magnon modes in the presence of electromagnetic fields.

In the numerical calculations we present below, we will treat all six modes described by Eqs.~(\ref{eq:qAF_Hyb}) and (\ref{eq:qa0}) on equal footing.
However, it can be shown that when $\omega_s/\omega_0\to 0$ or $\lambda_a\omega_0/c\to 0$, the four hybrid modes of Eq.~\eqref{eq:qAF_Hyb} decouple into pure polariton and magnon modes.
In this limit, the modes with $q = \pm q_{{\rm A}+}$ return to the polariton modes for $a=0$~\cite{Bludov:2DMat2019} with
\begin{align}
    q^2_{{\rm A}+} \to
    %q^2_{\text{AP}} =
    k^2+\frac{\omega^2}{c^2}\frac{\omega_0^2+2\omega_s^2-\omega^2}{\omega^2-\omega_0^2},
\end{align}
%where $\mu_{m,0}=\text{lim}_{\lambda_a\to0}\mu_{m}$, 
and the other two become magnonic modes with
\begin{align}
    q^2_{{\rm A}-} \to 
    %q^2_\text{AM}=
    k^2-\frac{1}{\lambda_a^2}\frac{\omega^2-\omega_0^2}{\omega_0^2}.
    \label{eq:qAM}
\end{align}
We will make the realistic assumptions that the domain-wall length in the AFI is sufficiently tiny, $\lambda_a k \ll \omega_s/\omega_0$, such that the four modes of Eq.~\eqref{eq:qAF_Hyb} almost fully decouple. 
In this limit, SMPs are mostly carried by localized polariton-like modes with $q=\pm q_{{\rm A}+}$. The remaining four modes become localized magnon-like modes with $q=\pm q_{{\rm A0}}$ and propagating magnon modes with $q=\pm q_{{\rm A}-}$.
This will allow for a perturbative analytic treatment of the SMP mode, where the contribution of the magnon(-like) modes can be assumed small. Furthermore, since the SMP has frequencies in the Restrahlen band $\omega_0<\omega<\sqrt{\omega_0^2+2\omega_s^2}$~\cite{Hale:PRB2022,Bludov:2DMat2019}, it follows that for real frequencies $\omega\in\mathcal{R}$, both $q^2_{{\rm A}+},q^2_{{\rm A}0}>0$ represent localized modes, while $q^2_{{\rm A}-}<0$ represent propagating magnon-like modes.

\emph{Bulk N--}
We first look at the semiconductor case without spin-Hall response $\theta_\text{SH}=0$, where Maxwell's equations yield two polariton modes,
\begin{align}
    q^2_\text{N--P}&=k^2-\frac{1}{c^2}\left(\epsilon_\infty\omega^2+\omega_p^2\frac{i\omega\tau}{1-i\omega\tau}\right),\label{eq:qsp}
\end{align}
and the spin-diffusion equation \eqref{eq:SpinDiffusionEq} yields two pairs of spin-accumulation modes with equal $q$, set by
\begin{equation}
    q_\text{N--SD}^2=k^2+\frac{1-i\tau_\text{sf}\omega}{\lambda_\text{sf}^2},\label{eq:qSMMu}
\end{equation}
where $\lambda_\text{sf}^2=D_0\tau_\text{sf}$ is the the spin-flip diffusion length. All solutions of Eqs.~(\ref{eq:qsp}) and (\ref{eq:qSMMu}) yield a $q$ with a non-zero real component and thus correspond to localized modes.

Next, we allow for a non-zero spin-Hall angle $\theta_\text{SH}$. Solving for spin accumulation as a function of the electric field $E_z$, we find $\bmu^s=\chi_\mu\nabla\times\bE$, with the following response function,
\begin{equation}
    \chi_\mu=-\frac{\sigma(\omega)\tau_{\mathrm{sf}}}{e\nu}\frac{\theta_\text{SH}}{1+\lambda_\mathrm{sf}^2(k^2-q^2)-i\tau_\mathrm{sf}\omega}.\label{eq:SMSpinAccSus}
\end{equation}
Since the spin accumulation gives rise to a charge current density [see Eqs.~(\ref{eq:SpinDispersionCurrent}) and (\ref{eq:SpinHallCharge})]
\begin{equation}
    \bj^\text{c,SH}=-\theta_\text{SH} D_0e\nu\nabla\times\bmu^s,
\end{equation}
the Maxwell's equations take the same form as in Eq.~\eqref{eq:Vacuum_qMat}, but replacing $\epsilon_0\to\epsilon_0\epsilon$ where
\begin{equation}
    \epsilon=\epsilon_\infty+\frac{i\sigma(\omega)}{\epsilon_0\omega}\Big[1+\theta_\text{SH}^2\frac{\lambda_\text{sf}^2(k^2-q^2)}{1+\lambda_\text{sf}^2(k^2-q^2)-i\tau_\text{sf}\omega}\Big].
\end{equation}
Solving these equations now yields four modes, with
\begin{align}
    q_{\text{N}\pm}^2= {} & {} \frac{\tilde{q}^2_{\text{N--P}}+q_{\text{N--D}}^2}{2}\label{eq:qsmpm}\\
    {} & {} \pm\sqrt{\Big(\frac{\tilde{q}^2_{\text{N--P}}-q_{\text{N--D}}^2}{2}\Big)^2-i\frac{\theta_\text{N--H}^2\omega\sigma(\omega)[1-i\tau_\text{sf}\omega]}{\epsilon_0c^2\lambda_\text{sf}^2}},\nonumber
\end{align}
where we introduced the notation
\begin{align}
    \tilde q^2_\text{N--P}&=k^2-\frac{1}{c^2}\left[\epsilon_\infty\omega^2+\left(1+\theta_\text{SH}^2\right)\omega_p^2\frac{i\omega\tau}{1-i\omega\tau}\right].
\end{align}
These four modes are hybrid, resulting from a mixing of the polariton modes corresponding to Eq.~(\ref{eq:qsp}) with two of the spin-accumulation modes with $q=\pm q_{\rm SD}$ described by Eq.~(\ref{eq:qSMMu}).
It is easy to see from Eq.~(\ref{eq:qsmpm}) that these modes indeed decouple again in the limit $\theta_{\rm SH}\to 0$.
The two remaining spin-accumulation modes, which also have $q = \pm q_{\rm N- SD}$, remain unaltered and do not couple to the electromagnetic field by satisfying $\nabla\times\bmu^s=0$. These modes have no electromagnetic field components and are purely composed of spin accumulation. They can be understood as superpositions of spin accumulations $\mu^{s}_{x}$ and $\mu^{s}_{y}$ polarized along $\ex$ and $\ey$ respectively, such that the resulting charge current from the spin-Hall effect cancels between the two polarization-components. 

\section{Include Boundary Conditions}\label{sec:bcs}
A non-trivial solution for the whole bilayer system must, in addition to satisfying the equations in the bulk, also satisfy all the boundary conditions given by Eqs.~\eqref{eq:MaxwellBC}, \eqref{eq:BCAll}, and \eqref{eq:BCvac}.
We thus look for a superposition of the bulk modes found in the previous section, which also conforms to the continuity of electromagnetic fields and spin currents at the interfaces.

We again refer to Fig.~\ref{fig:SysGeom}, which depicts the setup we consider:
a semi-infinite bilayer system in the $yz$ plane, consisting of an AFI layer of width $d_A$ and an N layer of width $d_N$, with a vacuum on either side of the bilayer. 
Since we are looking for localized modes, the fields in the vacuum must vanish at $x\to\pm\infty$, implying that on each side, only one of the two solutions of Eq.~\eqref{eq:qVacSol} can be used.
Furthermore, we note that for real $k,\omega$, a solution only exists for $k>\omega/c$, and we will focus on this regime in the following.

We find $14$ modes that obey the bulk equations of motion: six in the AFI layer, six in the N layer, and one vacuum mode on each side of the bilayer.
Maxwell's equations require continuity of the fields across all interfaces as described by Eq.~\eqref{eq:MaxwellBC}. However, from Maxwell's equation \eqref{eq:TE_SpatE} in the Fourier space, $ikE_z=i\omega B_x$, it follows that continuity of $B_x$ \eqref{eq:MaxwellBC_B_Orth} is satisfied if $E_z$ is continuous across the interface since $k$ and $\omega$ are assumed equal for all modes.
Thus, we get six boundary conditions from the continuity of $H_y$ and $E_z$ for each of the three interfaces.
The boundary conditions for the spin currents, given by Eqs.~(\ref{eq:BCAll}) and (\ref{eq:BCvac}), consist of 12 equations.
However, since the spin current can only be polarized in the $xy$ plane, we discard all equations involving $j^s_z$ and are thus left with eight spin-current boundary conditions.
In total, we thus find 14 boundary conditions for 14 modes.

We first need to relate the different fields for each mode to impose the boundary conditions.
Each mode, labeled by $k$, we can characterize by a single amplitude $\alpha_k$; the relative amplitudes of the different electromagnetic field components follow from equation \eqref{eq:Vacuum_qMat} and \eqref{eq:AF_qMat} for the vacuum and the AFI layer, while a similar equation can be found for the fields in the N by combining the Maxwell equations \eqref{eq:MaxwellTE} with the spin-diffusion equation \eqref{eq:SpinDiffusionEq}.
Note that the pure spin-accumulation mode in the N layer is unique, as it consists purely of spin accumulation, whose relative amplitudes are given by $\nabla\times\bmu^s=0$.

By plugging the fields corresponding to each of the $14$ modes into the boundary conditions, the resulting set of equations can compactly be written in matrix form as $A\boldsymbol\alpha^T=0$, where $\boldsymbol\alpha=(\alpha_k)$ is a vector of the amplitudes of the modes and $A$ is a $14\times14$ matrix. For a solution to exist, it is necessary to have $\det{(A)}=0$, from which the dispersion relation $\omega(k)$ is found. The relative ratio of the amplitude of the modes can then also be seen from the eigenvector corresponding to the zero eigenvalues.
Below, we present results from solving the determinant equation numerically through Newton's method.
We then compare those to approximate analytic expressions based on an expansion in small spin-mixing conductances, spin-Hall angle, domain-wall, and spin-flip lengths.

\section{Results}\label{sec:res}
This section presents our numerical and analytical results by the model outlined in Sec.~\ref{sec:Model} as explained in Secs.~\ref{sec:bulk} and \ref{sec:bcs}.

For concreteness, we use the material parameters of \ce{Cr2O3} for the AFI layer, which has a long domain-wall length scale of $100\,\mathrm{nm}$ \cite{Cr2O3_1}, with an AF resonance frequency of $\omega_0=1\,\mathrm{ThZ}$, the magnetic anisotropy field $\omega_a/\gamma \approx 0.7 \,\mathrm{T}$ and the sublattice saturation magnetization $\mu_0M_s\approx 0.286\,\mathrm{T}$, yielding $\omega_s\sim 25\,\mathrm{GHz}$ \cite{Cr2O3_2,Cr2O3_3, Cr2O3_4}.

For the N layer, we use material parameters of an $n$-doped \ce{GaN}, which has $\lambda_\text{sf}\sim 200\,\mathrm{nm}$ \cite{GaNLsf}, $\tau_\text{sf}=60\,\mathrm{ps}$ \cite{GaNTsf} and $\epsilon_\infty=5$ \cite{SM_DielectricValues}.
Furthermore, with an effective mass of $m^* = 0.24\,m_e$ \cite{SM_DielectricValues} and a very low electron charge density $n_e = 7.5 \times 10^{15}\,\mathrm{cm}^{-3}$, we find a density of states per spin of $\nu=7.6 \times 10^{43}\mathrm{J}^{-1}\mathrm{m}^{-3}$
and plasma frequency $\omega_p \approx 10\,\mathrm{THz}$.
We assume $g_r\sim 10^{17}\mathrm{m}^{-2}$ in \ce{GaN} and a spin-Hall angle of $\theta_\text{SH}=2 \times 10^{-3}$ \cite{GrGaN,AndoNature2011:SchottkySMC}, even though it is expected that the spin-Hall angle at the AF resonance-frequency will differ from the measured DC-value.

The bilayer heterostructure supports two SMP modes that satisfy the set of boundary conditions localized at the left and right interface of the AFI layer.
The number of magnon modes in the AFI layer inside the Reststrahlen band depends on the AFI layer width $d_A$, with more magnon modes for wider AFIs. Furthermore, since the AFI layer carries the SMPs, the properties of the SMPs are sensitive to $d_A$ in a qualitative way. However, reducing the N layer width $d_N$ mainly reduces N-induced frequency shifts but yields no qualitative change. Once $d_N$ becomes sufficiently small, the system behaves like the vacuum replaced the N.
We thus set $d_N = 20\,c/\omega_0$ in our numerical calculations, corresponding to $d_N = 300\,\mu$m, which is much longer than the decay length of the SMPs and approximates a semi-infinite semiconductor. We will focus on two distinct limits for the width of the AFI layer: the narrow limit $k d_A \ll 1$ and the wide limit $k d_A\gg1$.

\subsection{Numerical results}
\begin{figure}[t]
     \centering
        \includegraphics[width=8.6cm]{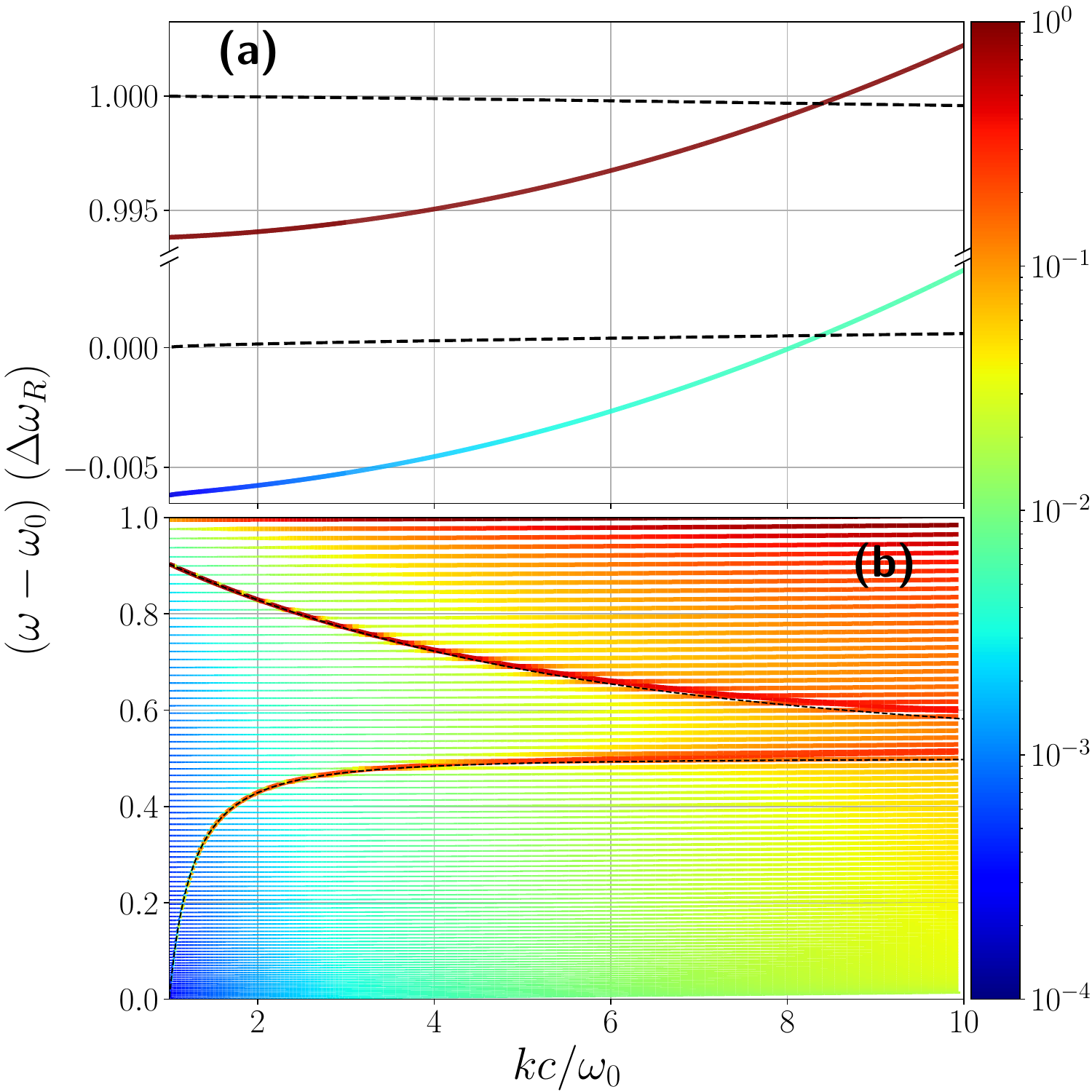}\\

      \caption{Dispersion magnon polaritons, both surface and bulk modes, for (a) a thin AFI with $d_A\omega_0 / c = 10^{-4}$ and
      (c) a wide AFI with $d_A\omega_0 / c = 3$, using the material parameters stated in the text.
      The color scale reflects the ratio of the magnetic field and magnetization in the AFI as given in equation \eqref{eq:color}. In (c), the redder lines are also thicker to aid readability.
      The black dashed lines show the case with zero exchange stiffness, spin pumping, and spin-Hall effect, i.e., the limit $\lambda_a,g_r,\theta_{\rm SH} = 0$.
      All frequencies are plotted in units of the width of the Reststrahlen band $\Delta \omega_{\rm R} = \sqrt{\omega_0^2 + 2\omega_s^2} - \omega_0$.}
    \label{fig:PolaritonDispFull}
\end{figure}
Figure \ref{fig:PolaritonDispFull} presents the dispersion of the two SMP modes from the two AFI interfaces in the narrow and wide limits.
The color code represents the ratio between the sum of the squared amplitudes of the magnetic fields and the in-plane magnetization of all the modes in the AFI layer,
\begin{equation}
    \text{color}=\frac{\sum_{k=\pm(q_{\text{A}\pm},q_{\text{A}0})}H_{k}^2}{\sum_{k=\pm(q_{\text{A}\pm},q_{\text{A}0})}M_{k}^2},\label{eq:color}
\end{equation}
where the blue corresponds to a mode carried by magnons and red to a mode carried equally by the electromagnetic field and magnons.
In the narrow limit, shown in Fig.~\ref{fig:PolaritonDispFull}(a), the two SMPs live close to the edges of the Reststrahlen band; the high-frequency SMP has a frequency close to $\omega\approx \sqrt{\omega_0^2+2\omega_s^2}$, while the low-frequency SMP has approximately $\omega\approx \omega_0$.
Figure \ref{fig:PolaritonDispFull}(b) shows the wide limit, where the SMP frequency lies between $\omega_0$ and $\sqrt{\omega_0^2+2\omega_s^2}$ for both modes, as shown by the dispersive curves.
The general behavior of these modes is very similar to that of the SMPs in a semi-infinite AFI--graphene heterostructure, see Ref.~\cite{Bludov:2DMat2019}.
The dense set of almost dispersionless states, absent in Fig.~\ref{fig:PolaritonDispFull}(a) as the AFI is too narrow, represents purely magnon-like states in the bulk of the AFI layer, carried by the propagating modes with $q_{\text{A}-}$.
They appear almost flat as their characteristic length scale $\lambda_a$ is much smaller than the polariton length scale $c/\omega_0$.
Due to their magnon-like nature, they mostly appear less red than the two polaritons for the same $k$ and $\omega$.
However, as $\omega\to\sqrt{\omega_0^2+2\omega_s^2}$ the contributions of the modes with $q_{\text{A}0}$ and $q_{\text{A}+}$ become significant. Since these two modes have significant electromagnetic field components, the weight of the magnetic field strength in the dispersionless states is increased in this limit; that is why they show up as increasingly red at higher frequencies.

The dispersion of the colored curves can also be seen to differ slightly from the black dashed line in figure \ref{fig:PolaritonDispFull}, representing the SMPs in the absence of exchange-stiffness, spin-pumping and spin-Hall angle. In Fig. \ref{fig:PolaritonDispFull}(b), the SMPs can also hybridize with the bulk magnonic states, forming anticrossings where two such states would otherwise have crossed.

\subsection{Analytic results}
To better understand the numerical results and how these modes are related to the microscopic details of the system, we perform analytical calculations for two limiting thicknesses of the AFI layer. We disentangle the effects of exchange stiffness, spin pumping, and the spin-Hall impact on the dispersion of the modes.
\subsubsection{{Thick AFI layer $kd_A \gg 1$}}
We start by considering the ``zeroth-order'' noninteracting case, i.e., without any magnon--polariton coupling, spin pumping, and spin-Hall effect.
Using the fact that for the systems of interest $\omega_s\ll \omega_0$, we find the following expressions for the high-frequency polaritons in the wide limit,
\begin{align}\label{HFP}
\begin{split}
    \omega_{\text{w}+}^{(0)} \approx {} & {} \omega_0+\frac{\omega_s^2}{2\omega_0}\left[\frac{2\left(c^2k^2+\omega_p^2-\epsilon_\infty\omega_0^2\right)+\omega_0^2}{\omega_p^2-(\epsilon_\infty-1)\omega_0^2}\right.\\
    {} & {} \left. -\frac{\sqrt{4c^2k^2\left(c^2k^2+\omega_p^2-\epsilon_\infty\omega_0^2\right)+\omega_0^4}}{\omega_p^2-(\epsilon_\infty-1)\omega_0^2}\,\right],\end{split}
\end{align}
and for the low-frequency polaritons,
\begin{equation}\label{LFP}
    \omega_{\text{w}-}^{(0)} \approx \omega_0+\frac{\omega_s^2}{2\omega_0}\left(1-\frac{\omega_0^2}{2c^2k^2-\omega_0^2}\right).
\end{equation}
In the limit of a high plasma frequency $\omega_p \gg \omega_0$, the high-frequency branch approaches $\sqrt{\omega_0^2+2\omega_s^2}$ at small $k$, while for large $k$ it goes to $\sqrt{\omega_0^2+\omega_s^2}$, consistent with the results presented before in Ref.~\cite{Bludov:2DMat2019}.
The low-frequency mode is localized at the AFI--vacuum interface and is thus insensitive to $\omega_p$ and $\epsilon_\infty$ in the thick film limit, which are properties of the N layer.
The mode starts at $\omega=\omega_0$ for $k=\omega_0/c$, and approaches $\sqrt{\omega_0^2+\omega_s^2}$ as $k\to\infty$, cf.\ Refs.~\cite{Bludov:2DMat2019,Hale:PRB2022}.

Next, we consider the leading-order corrections to these frequencies, Eqs. (\ref{HFP}) and (\ref{LFP}), that arise from coupling to the magnons, spin pumping, and spin-Hall effect.
We can construct an approximate model that describes the frequency shifts and the induced couplings between the polariton and magnon modes by expanding the governing equation ${\rm det}(A)$ as outlined in section \ref{sec:bcs},  in the relevant small parameters before solving for its zeros.
Apart from using $\omega_s/\omega_0\ll 1$ and $\omega_0/\omega_p \ll 1$, we can identify small parameters by comparing the length scales associated with the different types of modes that become mixed: polariton, magnon, and spin-accumulation modes.

The relevant length scale of the polaritons in the vacuum and the AFI layer is $\lambda_{\rm P} \sim k^{-1} \sim c/\omega_0$.
Using our assumption that $\lambda_a$ is small, we find from Eq.~(\ref{eq:qAM}) the typical length scale of the propagating magnons in the Reststrahlen band to be $\lambda_{\rm M} \sim \lambda_a \omega_0 / \omega_s$ for all $k$ of interest.
This suggests that we can treat $\lambda_{\rm M}/\lambda_{\rm P} \sim k\lambda_a\omega_0/\omega_s\ll 1$ as a small parameter in the expansion of ${\rm det}(A)$.
In the N layer, the polariton length scale becomes $\lambda_\text{SP} \sim (k^2+\omega_p^2/c^2)^{-1/2}$, while the spin-accumulation mode is characterized by the length scale $\lambda_\text{sf}$.
Thus, we can use $\lambda_\text{sf}/\lambda_{\text{SP}}\ll 1$ as a small parameter in the expansion.
Furthermore, we assume a small spin-Hall angle, so we only need to expand to the first order in $\theta_\text{SH}\ll 1$.
Finally, considering the effect of spin pumping perturbatively is slightly more complicated.
Part of the spin current pumped from the AFI layer into the N layer back-flow into the AFI layer, and for the material parameters that we use this back-flow spin current density is large compared to the bulk spin current density $j_\perp^{s,b}/j_\perp^{s,S}\sim g_\text{re}/\zeta_pq_{\text{S}\mu}\lambda_\text{sf} \gtrsim 1$; and thus the N layer does not act as a spin sink.
This back-flow being large yields the small dimensionless parameter,
\begin{equation}
\tilde g_e = \frac{\gamma_\text{re}}{1+\gamma_\text{re}/\zeta_pq_{\text{SD}}\lambda_\text{sf}}\ll 1,
\end{equation}
where $q_{\rm SD}$ is given by Eq.~(\ref{eq:qSMMu}), and we have introduced the following dimensionless parameters,
\begin{subequations}
    \begin{align}
        \gamma_\text{re}= {} & {} \frac{1}{2\pi}\sqrt{\frac{\gamma\tau_\text{sf}}{M_s\lambda_a\lambda_\text{sf}\nu}\frac{\omega_0}{\omega_a}}(g_r-\tilde{g}_r),\\
         \zeta_p= {} & {} \sqrt{\frac{\gamma\nu\hbar^2}{M_s\tau_\text{sf}}\frac{\omega_0}{\omega_a}\frac{\lambda_\text{sf}}{\lambda_a}}.
    \end{align}
\end{subequations}
$\gamma_\text{re}$ is an effective measure of the average strength of the bare spin pumping and back-flow, while the parameter $\zeta_p$ relates to the relative strength of the spin-pumping as viewed by the AFI and N layers.
One would find $\gamma_{\rm re} \approx 0.79$ and $\zeta_p = 1.3\times 10^{-3}$ for the explicit material parameters mentioned above.

The parameter $\tilde g_e$ can then be viewed as a renormalization of the effective spin-pumping parameter $\gamma_\text{re}$, as parts of the spin-current pumped into N back-flows back across the interface instead of diffusing into the N-layer. This then reduces the effective spin-pumping parameter, and below, we include spin pumping to the leading order in $\tilde g_e$. For the chosen parameters we find $\lvert\gamma_\text{re}/\zeta_pq_\text{N--SD}\lambda_\text{sf}\rvert\approx 77$, and as such the effective spin-mixing is significantly reduced compared to the bare value. This also means that further increasing the spin-mixing conductance $g_r$ does not increase the effect of spin-mixing, as the additional pumped spin-current back-flow into the AF.

Finite coupling to magnons gives rise to two effects: (i) a shift in the polariton frequency and (ii) hybridization of the polariton with the magnon, manifesting itself as anticrossings appearing wherever polariton and magnon modes intersect each other in the spectrum.

The spin pumping and spin-Hall phenomena present small additional contributions to both the frequency shift $\Delta\omega_\pm$ and the hybridization strengths.
In the following, we analyze all contributions perturbatively in terms of the small parameters introduced above.

\begin{figure}[t]
    \centering
    \includegraphics[width=8.6cm]{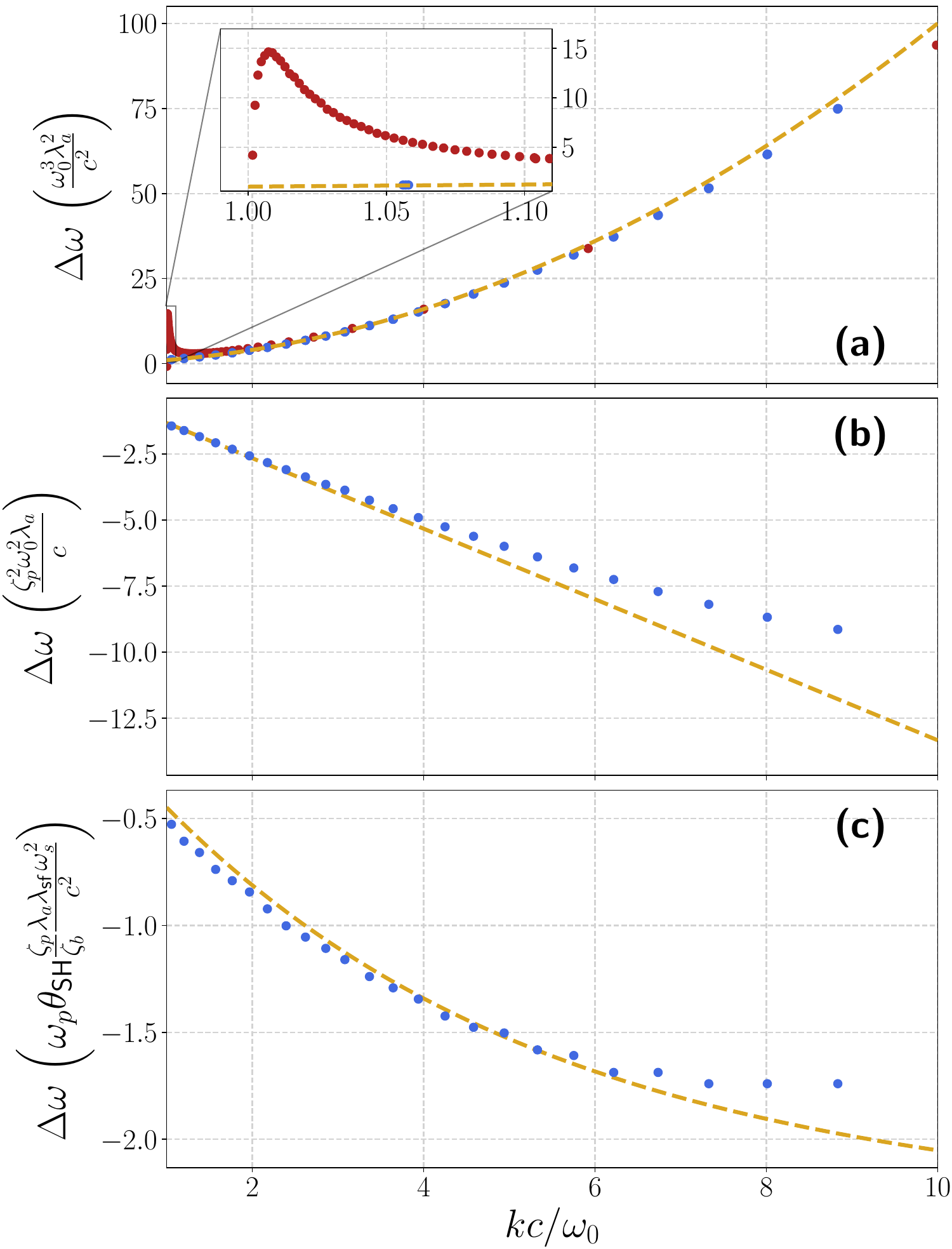}
    \caption{The change in the real part of $\omega(k)$ of the two polaritons for a wide AF with $d_A= 3.98 \, c/\omega_0 \approx 1.2\,$mm resulting from (a) including finite polariton--magnon mixing and, additionally, including (b) finite spin pumping across the AFI/N interface or (c) a spin-Hall effect in the N.
    The dots show the numerically calculated shift of the frequencies, where the red and blue dots in (a) distinguish the shift for the low- and high-frequency modes, respectively.
    %The numerical solution is given only where $k$-values equally far from points where the polariton hybridizes with the magnon is shown.
    As the text explains, the orange dashed lines show the analytically calculated corrections.
    For all plots, we used the material parameters stated in the text.}
        \label{fig:WideShiftsSMP}
\end{figure}

We start by analyzing the frequency shifts arising from the polariton--magnon coupling and setting $\tilde g_e, \theta_{\rm SH} = 0$. Later, we add the finite spin pumping and spin-Hall phenomena.
From Eq.~\eqref{eq:SpinCurrAF} and $\delta \bn=\chi_n\ez\times\bH$, we see that the spin current in the AFI layer scales with $\partial_x\bH=q\bH$, and thus it follows that if $\lambda_{\rm M}/\lambda_{\rm P} \sim k\lambda_a\omega_0/\omega_s\ll 1$, the polaritons only elicit a weak response from the magnons as the corresponding magnetic fields of the magnon modes are much smaller than for the polariton. In this limit, we find the leading-order correction in $\lambda_{\rm M}/\lambda_{\rm P}$ to the frequency shift as,
\begin{equation}
    \Delta\omega_{\text{w}\pm}^{\rm MP} = \left(k\lambda_a\right)^2\omega_0,\label{eq:betaMP}
\end{equation}
whose $k$-dependence are the same as that of a gapped AFI at long wavelengths \cite{Rezende}.
Figure~\ref{fig:WideShiftsSMP}(a) shows the frequency shift given by this expression using the material parameters given above (orange dashed line) compared to the numerically calculated shift of the two modes (colored dots), where the blue dots show the high-frequency polariton localized at the AF/N interface, while red dots are for the low-frequency polariton localized at the AFI--vacuum interface. The frequency is evaluated at $k$-values at the midpoint between two neighboring anticrossings to separate the shifts from the hybridization.
A small peak at small $k$ for the lower polariton mode (shown in the inset) is due to a strong hybridization with the mode with $q_{{\rm A}-}$:
From Eq.~\eqref{eq:qAM}, it follows that $q_{{\rm A}-}\to \pm k$ when $\omega\to\omega_0$, and thus the modes are no longer well separated. This shift can be understood as the additional dispersion resulting from the magnonic nature of the polariton since the magnons also disperse like $(k\lambda_a)^2/2$.

We then compute the effect of spin pumping by including a finite $\tilde g_e$, 
\begin{equation}
    \Delta\omega_{\text{w}+}^{\rm SP} = -\frac{i\tilde g_e\zeta_pk\lambda_a\omega_0}{4}.\label{eq:betasp}
\end{equation}
Since this correction is generally a complex frequency, it can not only induce a frequency shift but also introduce a damping of the mode. As the phase of $\tilde g_{e}$ is given by the factor $\sqrt{1-i\tau_\text{sf}\omega_0}$ stemming from $q_{\rm N-SD}$ when $g_\text{re}/\zeta_p\gg 1$, the frequency shift is only comparable to the damping if $\tau_\text{sf}\gtrsim\omega_0$.
Note that the lower-frequency polariton mode is unperturbed by the spin pumping $\Delta\omega_{\text{w}-}^{\rm SP}=0$, as it is localized at the AFI--vacuum interface. 
In Fig.~\ref{fig:WideShiftsSMP}(b), we compare the additional frequency shift of the high-frequency polariton predicted by Eq.~(\ref{eq:betasp}) (orange dashed line) with the numerically calculated shift (blue dots).
Since the effect of spin pumping on the polariton mode vanishes in the absence of polariton--magnon coupling, the correction Eq. (\ref{eq:betasp}) is accompanied by the correction Eq. (\ref{eq:betaMP}).
The numerically calculated shift in Fig.~\ref{fig:WideShiftsSMP}(b) is thus found as the difference in frequency between the case with a finite spin pumping and polariton--magnon coupling and the case with only polariton--magnon coupling.

We finally compute the effect of a finite spin-Hall mixing, 
\begin{align}
\begin{split}\Delta\omega_{\text{w}+}^\text{SH}=-&\theta_{\rm SH}k\lambda_a\frac{\tilde g_e}{\zeta_b} \frac{\omega_s^2}{2\omega_0}\frac{\lambda_\text{sf}\sqrt{k^2+\frac{\omega_p^2}{c^2}}}{\sqrt{1-i\tau_\text{sf}\omega_0}}\\
    &\times\left(1-\frac{k}{\sqrt{k^2+\frac{\omega_p^2}{c^2}}}\right),
    \end{split}\label{eq:betaSH}
\end{align}
where we introduced the following dimensionless parameter,
\begin{equation}
    \zeta_b = \frac{\omega_p^2}{\omega_0^2}\sqrt{\tau_\text{sf}\omega_0}\sqrt{\frac{\gamma M_s\omega_a}{c^4e^2\nu}\frac{\lambda_a}{\lambda_\text{sf}}},
\end{equation}
which characterizes the relative ratio of how strongly the two layers are influenced by the spin transferred from the other layer.
We find $\zeta_b=9.7\times 10^{-5}$ for the material parameters used in this article.
The correction described by Eq. (\ref{eq:betaSH}) originates from the spin current pumped from the AFI layer into the N layer, being converted into an electric and magnetic field in the N layer through the spin-Hall effect, and is thus again only affecting the high-frequency mode, i.e., $\Delta\omega_{\text{w}-}^\text{SH}=0$.
There is also a correction to the frequency proportional to $\zeta_b$ for the opposite ``loop'' caused by the back-flow spin current that converts spin accumulation in the N layer into electromagnetic fields in the AFI layer.
This correction, however, is much smaller than $\Delta\omega_+^\text{SH}$ since $\zeta_b\ll 1$. 
In Fig.~\ref{fig:WideShiftsSMP}(c), we compare the analytically found correction following from Eq. (\ref{eq:betaSH}) (orange dashed curve) with the numerically calculated frequency shift of the mode (blue dots), again focusing on the difference between the case with only polariton--magnon coupling and the case with additional spin-Hall mixing.

The total approximate expression for the frequency of the two polariton modes in the thick limit of the AFI layer, in the presence of all mechanisms considered here, thus becomes,
\begin{equation}
    \Delta \omega_\pm \approx \omega_{{\rm w}\pm}^{(0)} + \Delta \omega_{{\rm w}\pm}^{\rm MP} + \Delta \omega_{{\rm w}\pm}^{\rm SP} + \Delta \omega_{{\rm w}\pm}^{\rm SH}.
\end{equation}
From Fig.~\ref{fig:PolaritonDispFull}(b), it can be seen that the upper polariton coupled to N exhibits a negative group velocity, and as shown in Fig.~\ref{fig:WideShiftsSMP}(b,c), this effect can in principle be enhanced by spin mixing and SHE. However, the effect is fragile due to the small spin-Hall angle in most semiconductors and the reduced effective spin-mixing due to a large back-flow spin-current from the N, reducing the effective spin-pumping. However, this large back-flow is also responsible for the pure spin-mixing induced frequency shift in the absence of SHE as given in Eq. \eqref{eq:betasp}; in its absence, the spin-mixing would only lead to an increased damping of the polariton. On the flip side, the coupling to the magnonic modes in the AF imparts a weak magnonic dispersion on the surface polariton, reducing the negative group velocity. As such, the combined effect will be very material-dependent.

We now turn to the hybridization of the polaritons with the dispersionless magnon-like states.
To investigate the hybridization strength, we calculate the size of the avoided crossing gaps, i.e., the minimal difference in frequency between two given states as a function of $k$ numerically and analytically.
To construct an analytical model, we make use of the same approximations as for the frequency shifts, expanding the determinant ${\rm det}(A)$ from section \ref{sec:bcs} in the small parameters mentioned above.

\begin{figure}
     \centering
    \includegraphics[width=0.44\textwidth]{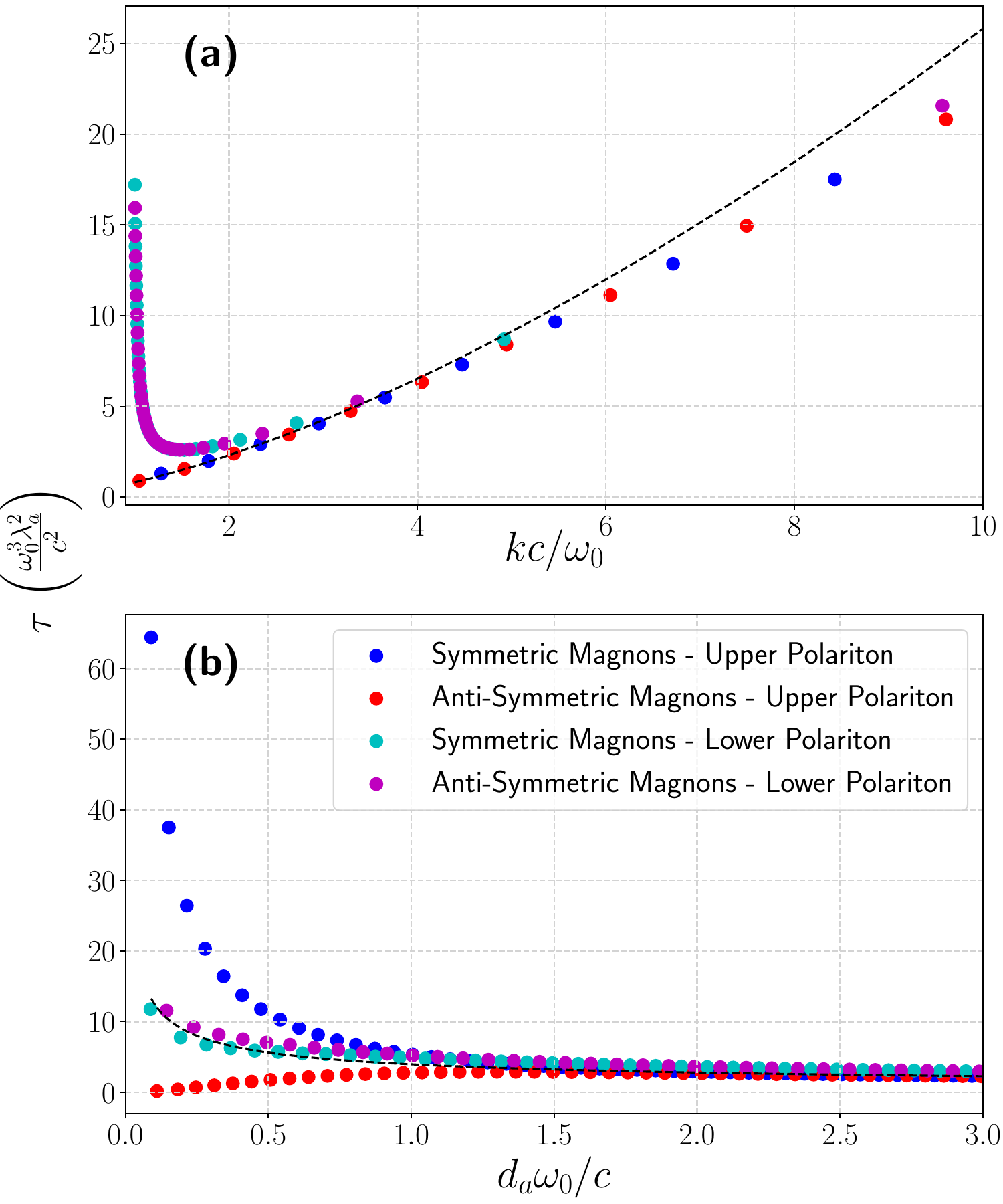}
      \caption{The size of the anticrossings between the high-frequency polariton mode and the magnon modes, given by the approximate expression (\ref{eq:tau}) (black dashed curves) and calculated numerically (colored dots).
      The blue dots show the anticrossings with symmetric magnon modes and the red dots with antisymmetric modes.
      (a) The splitting as a function of $k$ for $d_A=3\, c/w_0$.
      (b) The splitting $\tau^\text{MP}_{{\rm w}\pm}$ as a function of $d_A$, for $k=2\,\omega_0/c$.
      In the numerical calculations we used $\theta_\text{SH},\gamma_\text{re}=0$ throughout.}
        \label{fig:GapMag}
\end{figure}

Here, we focus only on the contribution from the polariton-magnon coupling and do not consider the contributions of the spin mixing and spin-Hall phenomena; the latter are minor corrections.
In the wide limit $kd_A \gg 1$, we find the size of the avoided crossing gaps up to the leading order in $\lambda_M/\lambda_P$
\begin{equation}
\delta^\text{MP}_{{\rm w}\pm} = k^2\lambda_a^2\omega_0\sqrt{\frac{2}{kd_A}}.\label{eq:tau}
\end{equation}
Importantly, this means that the width of the anticrossings decreases as the width of the AFI layer increases while increasing with increasing $k$.
In Fig.~\ref{fig:GapMag}, we compare this analytical expression (black dashed curves) with the numerically found splittings (colored dots) of the anticrossings between the polariton mode and the magnon mode.
Fig.~\ref{fig:GapMag}(a) explores the hybridization as a function of $d_A$ and Fig.~\ref{fig:GapMag}(b) as a function of $k$.
The approximate result captures the behavior very well for thick AFI layers. The sharp increase as $k$ approaches $\omega_0/c$ for the lower polariton is due to the same effect as the peak in Fig~\ref{fig:WideShiftsSMP} (a), namely that the separation between the magnonic $q_{\text{A}-}$-mode and the polariton diminishes as $q_{\text{A}-}\to k$ in this limit.

The two ``branches'' of splittings appearing at smaller $d_A$ can be understood as follows:
For a thinner AFI layer, the two polariton modes are no longer well separated at the two interfaces of the AFI layer.
The high-frequency mode ultimately becomes a mode with symmetric $E_z$-field along the $x$--direction, and the low-frequency mode becomes an antisymmetric mode.
For the high-frequency case, one thus expects the hybridization with antisymmetric magnon modes to vanish for a thin AFI layer (red dots) and with symmetric modes to become stronger (blue dots).
The low-frequency mode displays a much weaker branching but is still visible, split at smaller $d_A$, but with the roles of the symmetric and antisymmetric magnon modes interchanged.

\subsubsection{Thin AFI layer $kd_A \ll 1$}
In this limit, the decay length of the polariton modes in the AFI is longer than the thickness of the AFI. The SMPs will no longer be located at the two separate interfaces but instead mix to form superpositions of the two $\pm q_{\text{A}+}$-modes decaying in opposite directions.
The high-frequency mode becomes a symmetric superposition of the two modes when viewed in terms of the electric field $E_z$, while the low-frequency polariton approaches an antisymmetric superposition.
The frequency of the upper polariton in the uncoupled limit $\lambda_a,\gamma_\text{re},\theta_\text{SH}\to0$ approaches $\omega_{\text{n}+}^{(0)}\approx\sqrt{\omega_0^2+2\omega_s^2}$, while the lower polariton approaches $\omega_{\text{n}+}^{(0)}\approx \omega_0$, see again Fig.~\ref{fig:PolaritonDispFull}(a,b). Notably in this limit $q_{\text{A}-}, q_{\text{A}0}\to k$ for the lower and upper mode, respectively.

We derive approximate expressions for the induced frequency shifts due to the polariton--magnon mixing, spin pumping, and the spin-Hall effect in the thin limit of the AFI layer.

\begin{figure}[t]
     \centering
        \includegraphics[width=0.44\textwidth]{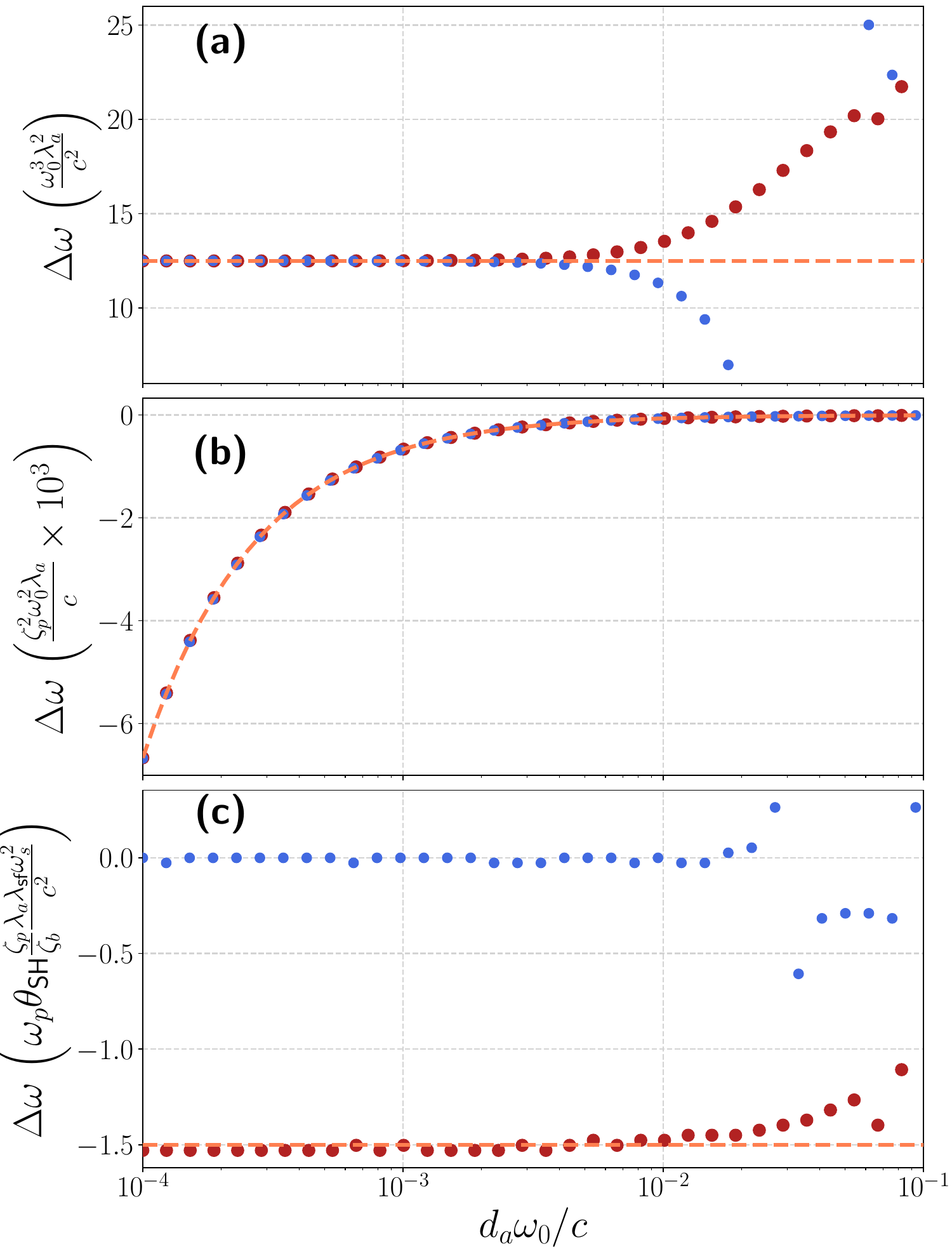}
      \caption{
      The change in the real part of $\omega(k)$ of the two polaritons for a narrow AF as a function of $d_A$ for a fixed $k=5\,\omega_0/c$ resulting from (a) including finite polariton--magnon mixing and, additionally, including (b) finite spin pumping across the AF/N interface or (c) a spin-Hall effect in the N.
      The dots show the numerically calculated shifts for the high-frequency polariton (blue) and the low-frequency polariton (red).
      The seeming divergence at certain larger $d_A$ results from anti-crossings with magnon-like states in the AF.
      The orange dashed curves show the approximate analytical results derived for the narrow limit.}
        \label{fig:DaDispFull}        
\end{figure}
The frequency shift due to coupling with the magnons, to the lowest order in $\lambda_\text{M}/\lambda_\text{P}$, becomes,
\begin{align}
    \Delta\omega_{\text{n}\pm}^\text{MP} = \frac{1}{2}k^2\lambda_a^2\omega_0.
\end{align}
In Fig.~\ref{fig:DaDispFull}(a), we compare this analytical result (orange dashed line) with the numerically calculated shifts (colored dots), where the blue (red) dots correspond to the (high-) low-frequency mode. Note that the perceived divergence as $d_A$ is due to the hybridization with magnon modes, which, as seen from figure \ref{fig:GapMag}(a), is strong for a thin AFI layer.

The effect of additionally adding the spin pumping leads then to a frequency change of
\begin{align}
    &\Delta\omega_{\text{n}\pm}^\text{sp}=-\frac{ig_e\zeta_p\lambda_a\omega_0}{8d_A}.
\end{align}
Just as in the thick limit, this is a complex frequency that leads to a frequency shift and damping to the mode.
We show this result in Fig.~\ref{fig:DaDispFull}(b) (orange dashed line), again comparing it to the numerically calculated shift of the high-frequency (blue dots) and low-frequency (red dots) modes.
This correction diverges in the limit $d_A\to0$, which is reasonable. For a very thin AFI layer, the total magnetization is small and thus more quickly pumped into the N layer, where it decays. Note, however, that the frequency shift is constant as a function of $k$ and thus does not change the group velocity of the polariton mode. 

Finally, the effect of a finite spin-Hall mixing yields,
\begin{equation}
\Delta\omega_{\text{n}-}^\text{SH}=\Delta\omega_{\text{w}+}^\text{SH}\sqrt{1-\frac{\omega_0^2}{c^2k^2}},\label{eq:narrowSH}
\end{equation}
for the low-frequency mode, whereas for the high-frequency mode $\Delta\omega_{\text{n}+}^\text{SH}\approx0$.
These results are illustrated in Fig.~\ref{fig:DaDispFull}(c), where the orange dashed line indicates $\Delta\omega_{\text{n}-}^\text{SH}$ and the blue(red) dots show the numerically calculated frequency shift of the (high-)low-frequency polariton.
The strongly reduced shift for the lower mode follows from the symmetry of the polaritons, which hybridize in the thin film limit into modes with either symmetric or anti-symmetric fields in the AFI layer. Due to the boundary conditions on the electromagnetic fields at the interfaces with the vacuum and N layer, the ratio of $H_y/E_z$ must remain finite even in the thin film limit. As the frequency of the high-frequency polariton approaches $\omega\sim\sqrt{\omega_0^2+2\omega_s^2}$ in the thin limit, $q_{\text{AF}+}\sim k$ remains finite, and the anti-symmetric fields will be expected to be much smaller than their symmetric counterparts by a factor $\sinh(q_{\text{A}+}d_A)/\cosh(q_{\text{A}+}d_A)\sim q_{\text{A}+}d_A\ll1$. Furthermore, as $|\mu_m(q_\text{A+})|\ll 1$ for $\omega\sim\sqrt{\omega_0^2+2\omega_s^2}$, it follows from Eq.~\eqref{eq:MaxwellTE} that only for a symmetric $E_z$-field does the ratio $H_y/E_z\propto \tanh[q_\text{AF+}d_A/2]/\mu_{m}(q_{\text{A}+})$ not diverge at the interface. The high-frequency mode must therefore have a symmetric $E_z$- and $H_x$-field, and an anti-symmetric $H_y$-field, and thus Eq.~\eqref{eq:MaxwellTE} implies $H_x\gg H_y$. 
The low-frequency polariton will, in turn, have the opposite field symmetries. However, deriving this result is slightly more complicated as both $\mu_{m,\text{A}+}$ and $q_{\text{A}+}$ diverges as $\omega\to\omega_0$.
From Eq.~\eqref{eq:SpinCurrAF}, where the spatially varying component of $\bn$ is given by $\delta \bn\propto\ez\times\bH$, it follows that the $H_i$-field mostly pumps spin into the $\mu_i^s$ component. However, as $\lambda_\text{sf}\ll c/\omega_0$, the spin current in the N layer flows more strongly along the $x$ direction, and thus $\mu_y^s$ couples much more strongly than $\mu_x^s$ to the electric polarization through the spin-Hall effect. Therefore, the small $ H_y$ field in the high-frequency polariton is why the spin-Hall-induced frequency shift is much weaker than in the lower mode.
As $d_A$ increases, the lower mode becomes more localized towards the left interface, and thus, the coupling to the N layer vanishes. At the same time, the upper mode localizes at the right interface, which causes the (anti)-symmetry of the fields to be lost, thus causing the reappearance of a finite frequency shift.

\section{\label{sec:Conclusion}Conclusion and Summary}

We have developed a formalism for the study of surface magnons polaritons in an antiferromagnetic insulator coupled to a semiconductor. Starting from Maxwell's equations combined with the Landau-Lifshitz-Gilbert equation for the dynamical electric and magnetic fields, we generalized this conventional description with spin-pumping, spin torque, spin Hall effects, and inverse spin Hall effects. We computed numerically the resulting dispersion relations for the eigen-excitations in the hybrid system. We also found analytical expressions for the influence of spin-pumping and spin Hall effects in the perturbative limits of thick and thin antiferromagnetic layers. There is a good agreement between the numerical and analytical results in appropriate limits. We found that conventional electromagnetic couplings are robust and typically dominate the dispersion relations, but spin-pumping and spin Hall effects give smaller frequency shifts in the perturbative regimes. Our analytical results show how material parameters control the impact of spin-pumping and spin-transfer torques on the surface magnon polaritons and could be used as a guide to explore new materials for magneto-plasmonic applications.

\section*{Acknowledgements}
The Research Council of Norway has supported this work through its Centers of Excellence funding scheme, project number 262633, "QuSpin."

\bibliography{references}

\end{document}